# Process Verification of Magnetic Ion Embedded Nanodiamonds Using Secondary Ion Mass Spectroscopy

Bo-Rong Lin[1,2], Chien-Hsu Chen[2], Srinivasu Kunuku[2], Tung-Yuan Hsiao[2], Hung-Kai Yu[2], Tzung-Yuang Chen[3], Yu-Jen Chang[4], Li-Chuan Liao[4], Chun-Hsiang Chang[5,6], Fang-Hsin Chen[6,7,8], Huan Niu[2*] and Chien-Ping Lee[1]

[1] *Department of Electronics Engineering and Institute of Electronics, National Chiao Tung University, Hsinchu, Taiwan*

[2] *Accelerator Laboratory, Nuclear Science and Technology Development Center, National Tsing Hua University, Hsinchu, Taiwan*

[3] *Health Physics Division, Nuclear Science and Technology Development Center, National Tsing Hua University, Hsinchu, Taiwan*

[4] *Bioresource Collection and Research Center, Food Industry Research and Development Institute, Hsinchu, Taiwan*

[5] *Department of Biomedical Engineering and Environmental Sciences, National Tsing Hua University, Hsinchu, Taiwan*

[6] *Radiation Biology Research Center, Institute for Radiological Research, Chang Gung University/Chang Gung Memorial Hospital, Taoyuan, Taiwan*

[7] *Department of Radiation Oncology, Chang Gung Memorial Hospital, Taoyuan, Taiwan*

[8] *Department of Medical Imaging and Radiological Sciences, Chang Gung University, Taoyuan, Taiwan*

## Abstract

Ion implantation is used to create magnetic ion embedded nanodiamonds for use in a wide range of biological and medical applications; however, the effectiveness of this process depends heavily on separating magnetic nanodiamonds from non-magnetic ones. In this study, we use secondary ion mass spectrometry to assess the distribution of magnetic ions and verify the success of separation. When applied to a series of iron/manganese embedded nanodiamonds, the sorting tool used in this study proved highly effective in selecting magnetic nanodiamonds. This paper also discusses the major challenges involved in the further development of this technology.

* Electronic mail: **hniu@mx.nthu.edu.tw**

## I. INTRODUCTION

Materials science has been revolutionized by the development of nanomaterials, which provide functional performance far exceeding that of conventional materials[1]. The adoption of nanomaterials has been particularly rapid in the field of medicine[2]. One such material is nanodiamonds (NDs), which provide surface functionalization with excellent bio-compatibility[3]. In our recent works, we developed a series of novel processes to fabricate magnetic ion embedded NDs[4]. It has been reported that fluorescent iron embedded NDs could be used as *in vivo* tracers with dual MRI/optical functionality[5,6]. We also fabricated fluorescent $^{10}$B embedded NDs as potential boron delivery agent for boron neutron capture therapy[7].

When using ion implantation to insert magnetic ions into the diamond lattice, there is little to no risk of magnetic ion release, due to the stability of the implanted ions[8]. However, the effectiveness of this process depends heavily on separating magnetic NDs from non-magnetic ones. The magnetic activated cell sorting (MACS) tool used in this study proved highly effective in collecting NDs with embedded magnetic ions; however, quantitative or qualitative measurements are required to verify the accuracy and precision of the selection process. In this study, we used secondary ion mass spectrometry (SIMS) to measure the distribution of magnetic ions within samples. Experiment results demonstrate the efficacy of the MACS tool in the selection of the magnetic ion embedded NDs.

## II. SAMPLES AND EXPERIMENTS

Fig. 1 presents the overall fabrication process. Monocrystalline ND powder (MSY 0-0.1 *um*) from Microdiamant in Switzerland was mixed in DI water for the coating of silicon wafers. The median size of the NDs was 50 *nm*, and the median tolerance was 0.04~0.06 *um*. After drying, the wafers were placed in a high-vacuum implantation chamber. The Ion Implanter (High Voltage Engineering Europe) was used for the implantation of iron and manganese ions. Iron ions were implanted using energy of 150 *keV* at a dose of $5 \times 10^{15}$ *ions/cm²*. Manganese ions were implanted using energy of 80 *keV* at a dose of $5 \times 10^{15}$ *ions/cm²*. Following ion implantation, the wafers were subjected to ultrasonic treatment using DI water to obtain a mixture comprising magnetic ion embedded NDs as well as normal (non-embedded) NDs (lower left of Fig. 1). A magnetic activated cell sorting tool (MACS Separator, Miltenyi Biotec Co.) was used as a filter for the collection of magnetic ion embedded NDs (the upper right of Fig. 1). After filtering, the sorting tool was removed from the magnetic stage to wash out the magnetic ion embedded NDs using DI water (the lower right of Fig. 1). The DI water was then removed via evaporation, leaving behind magnetic ion embedded NDs for use in subsequent experiments and analysis. A time-of-flight secondary ion mass spectrometer (ION-TOF, TOF-SIMS V) was used to measure the distribution of ions within the samples.

## III. RESULTS AND DISCUSSION

In this study, two types of magnetic ion embedded NDs were fabricated for investigation: iron embedded NDs (Fe-NDs) and manganese embedded NDs (Mn-NDs). Samples were prepared in four forms to verify the efficacy of magnetic separation. The first form comprised Fe-NDs or Mn-NDs on a silicon wafer immediately after removal from the implantation chamber. The second form comprised Fe-NDs or Mn-NDs filtered out of the initial mixture and then re-deposited on a silicon wafer. The third form was from the filtrate re-deposited on a silicon wafer (i.e., following the removal of magnetic ion embedded NDs). The fourth form was the wafer used for iron or manganese ion implantation (following the removal of coated NDs). The four types of sample were analyzed to verify the efficacy of the magnetic separation scheme used in the proposed fabrication processes.

Fig. 2 presents carbon and manganese SIMS profiles of the four samples based on Fe-NDs. Fig. 2 (a) presents the SIMS profile from as-implanted Fe-NDs, revealing a peak in the iron ion distribution (at 124 *nm*) at the top of the NDs layer. The distribution curve is similar to that obtained after ion implantation on bulk material. Fig. 2 (b) presents the SIMS profile from Fe-NDs filtered out of the original mixture. Following separation, the area tested by SIMS presented a uniform distribution of Fe-NDs, as expected. Fig. 2 (c) presents the SIMS profile from the filtrate re-deposited on silicon. Following separation, the area tested by SIMS should comprise normal NDs, such that the iron content is very low. In our analysis, the signal intensity of iron ions within the measured thickness was weak, indicating that the sample comprised mainly normal NDs. Fig. 2 (d) presents the SIMS profile from the silicon wafer used for iron ion implantation following the removal of NDs. The signal intensity of iron within the measured thickness was weak, indicating that the ND coating was of sufficient thickness to ensure that all high-energy iron ions were captured during ion implantation. An excessive quantity of NDs could hinder the separation process (i.e., increase the separation time); therefore, in the fabrication of Mn-NDs, we reduced the thickness of the initial ND coating.

Fig. 3 presents carbon and manganese SIMS profiles of the four samples based on Mn-NDs. Fig. 3 (a) presents the SIMS profile obtained from as-implanted Mn-NDs, showing an Mn ion distribution peak (at 55 *nm*) at the top of the NDs layer. The peak from Mn-NDs was shallower than that of Fe-NDs, due to the lower energy of the Mn ions. Fig. 3 (b) presents the SIMS profile from Mn-NDs filtered out of the initial mixture. As with the Fe-NDs, the area tested by SIMS presented a uniform distribution of Mn-NDs. The distribution of Mn ions again matched our expectations. Fig. 3 (c) presents the SIMS profile from the filtrate re-deposited on silicon (i.e., following the removal of Mn-NDs). In our analysis, the signal intensity of Mn ions within the measured thickness was again weak, indicating that the sample comprised mainly normal NDs. Fig. 3 (d) presents the SIMS profile from the silicon wafer used for manganese ion implantation (following the removal of Mn-NDs). We detected a signal associated with Mn within the silicon wafer. This could be attributed to the fact that we reduced the thickness of ND coating on the initial substrate. Nonetheless, the presence of manganese in the wafer would in no way have a negative impact on the fabrication process. In fact, reducing the thickness of the NDs coating could reduce the separation time by a factor of ten.

Due to similarities in preparation conditions, the SIMS signal intensity was comparable in all samples based on Fe-NDs and Mn-NDs. For the Fe-ND samples, the signal intensity at 1000 *nm* in sample 2 (Fig. 2 (b)) was 750x stronger than in sample 3 (Fig. 2 (c)). For the Mn-ND samples, the signal intensity at 1000 *nm* was 67x stronger in sample 2 (Fig. 3 (b)) than in sample 3 (Fig. 3 (c)). These results indicate the high selectivity of the sorting tool in the selection of NDs with embedded magnetic ions. Compared to a conventional magnetic separation machine, the tool featured in this study is easy to use (no complex setup) and highly economical.

## IV. DISCUSSION

In this study, we fabricated two types of magnetic ion embedded NDs (Fe-NDs and Mn-NDs) for analysis. Combining the context of this study and the experiences in the application of magnetic ion embedded NDs to biological applications[5-7], several directions are suggested for future improvement.

First, the SIMS results in this work make it clear that very few magnetic ion embedded NDs are available with single energy ion implantation. We suggest the use of multi energy ion implantation to vary the range of embedding and thereby increase the fabrication yield.

Second, the thickness of the initial ND coating is strongly correlated with the subsequent filtering time. Unfortunately, we found that it is difficult to control the thickness of the initial ND coating in its current form. It is possible that mixing NDs with photoresist would make it easier to obtain an ND coating of precise thickness. Further experiments will be required to verify this supposition.

Third, the effectiveness of nanoparticles for specific applications is largely size-dependent. At present however, the optimal size for NDs in biological applications has yet to be elucidated. Furthermore, we are still attempting to determine the optimal size for NDs in the proposed fabrication process.

Fourth, the effectiveness of magnetic NDs in biological applications is proportional to the number of embedded magnetic ions. High dose ion implantation is an obvious approach to achieving this goal; however, this could seriously damage the structure of the NDs. Researchers have yet to elucidate the relationship between energy, dose, and the structure of the resulting NDs. Calculating the number of magnetic atoms incorporated within an ND is another serious challenge.

## V. CONCLUSION

This paper reports on the SIMS measurements of iron embedded NDs and manganese embedded NDs. Our results demonstrate the high efficiency and specificity of the sorting tool used in this work for the selection of magnetic NDs. In practical experiments, most of the magnetic ion embedded NDs were obtained, and the discarded NDs were shown to contain very few magnetic ions. This paper also discusses some of the challenges involved in the further development of this technology. Despite these challenges, magnetic ion embedded NDs show considerable potential for a wide range of biomedical applications.

**Acknowledgement**

The authors acknowledge Team Union Ltd. for their financial supporting.

**Figure caption**

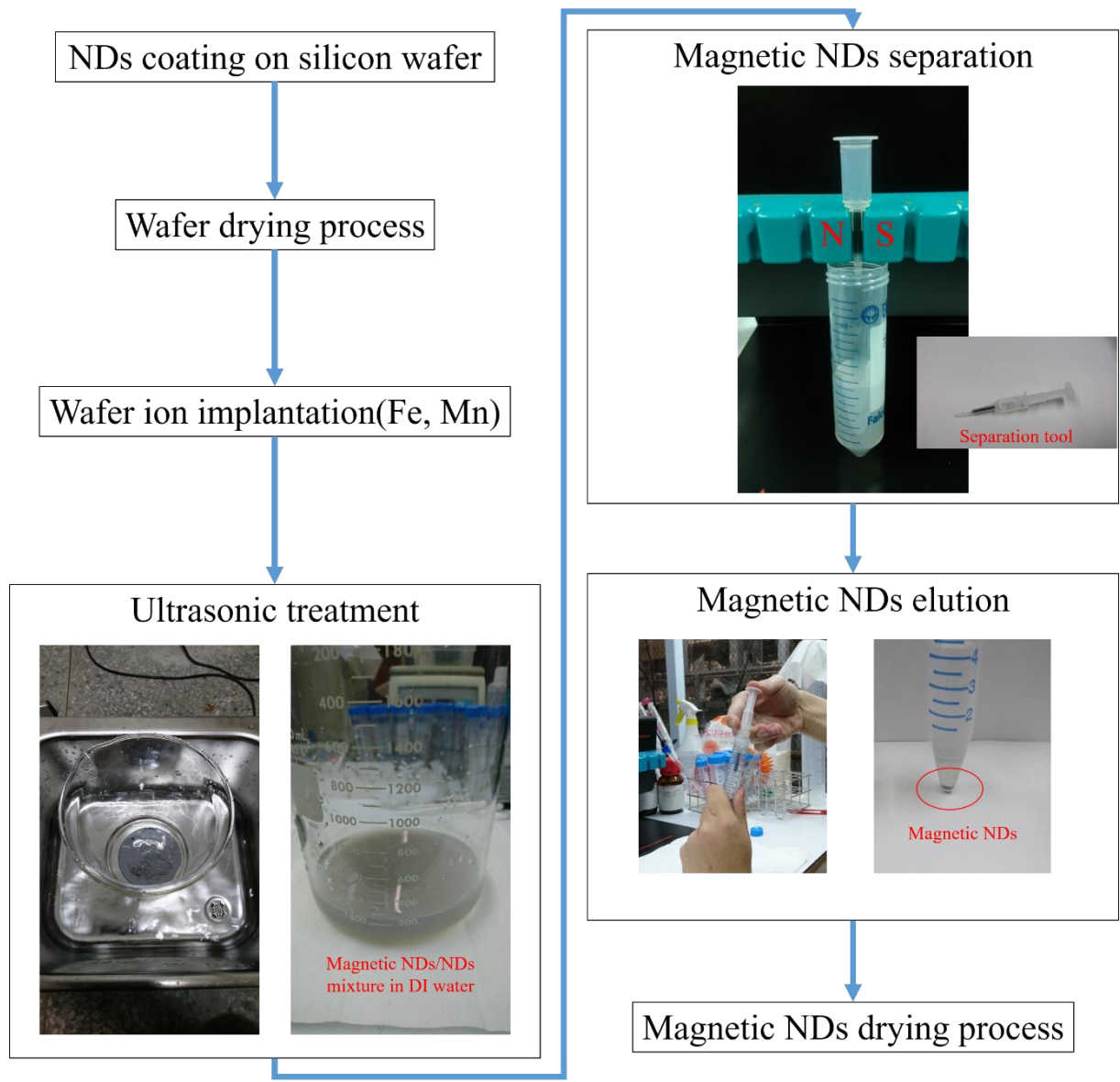

Fig. 1. Fabrication process flow of magnetic ion embedded NDs.

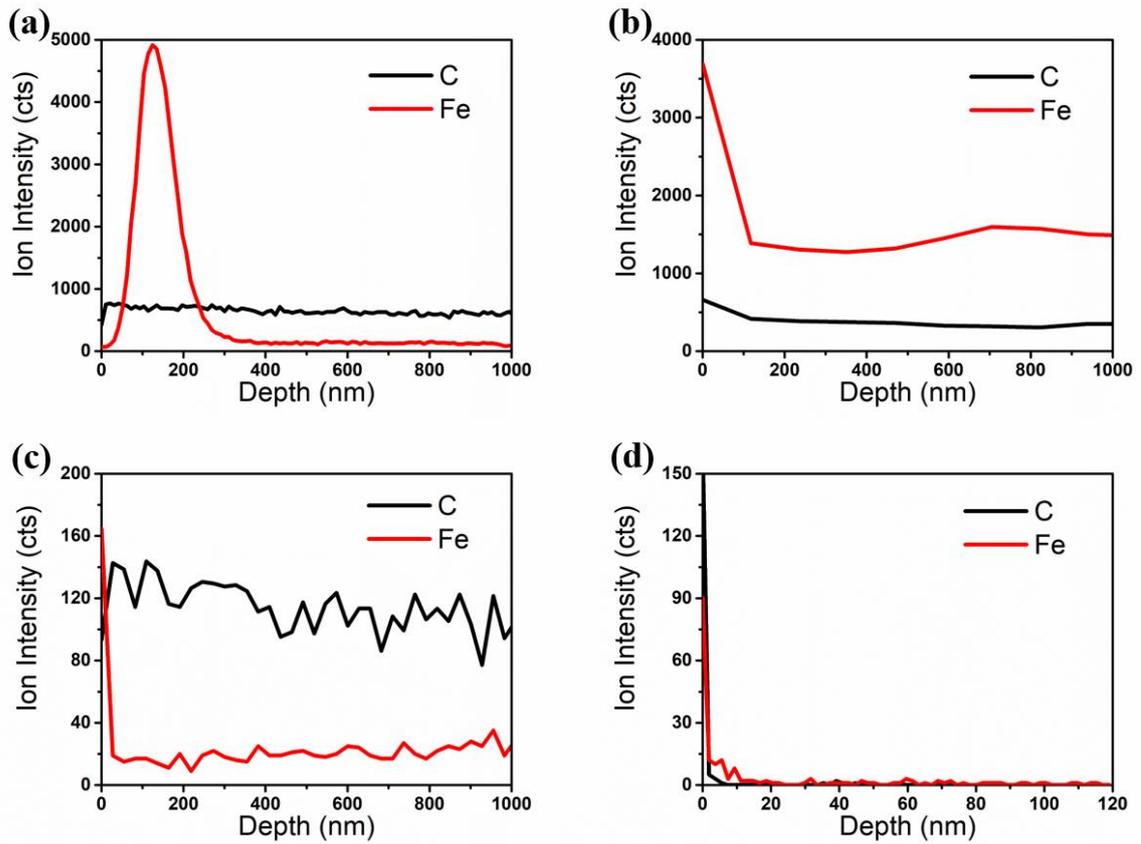

Fig. 2. Carbon and iron SIMS profile of (a) Fe-NDs as-implanted on silicon wafer, (b) filtered Fe-NDs re-deposited on silicon wafer, (c) filtrate re-deposited on silicon wafer, (d) silicon wafer used for iron ion implantation (following removal of NDs)

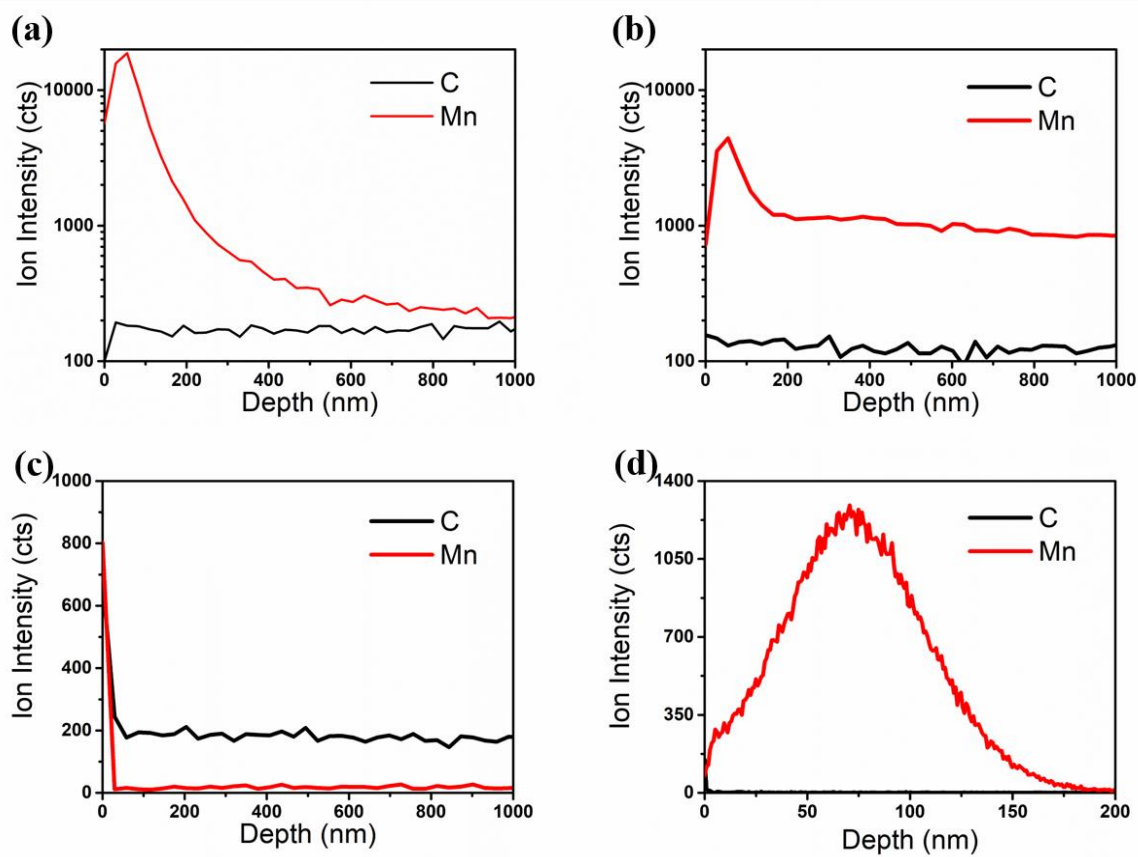

Fig. 3. Carbon and manganese SIMS profile of (a) Mn-NDs as-implanted on silicon wafer, (b) filtered Mn-NDs re-deposited on silicon wafer, (c) filtrate re-deposited on silicon wafer, (d) silicon wafer used for manganese ion implantation (following removal of NDs)